\documentclass[titlepage,preprint,nofootinbib,prd,tightenlines,byrevtex,amsmath]{revtex4}
\usepackage[pctex32]{graphicx}
\usepackage{amsmath}

\newcommand{\beq}{\begin{equation}}
\newcommand{\eeq}{\end{equation}}
\newcommand{\eq}[1]{eq.(\ref{#1})}

\begin{document}
\title {Three-Loop Radiative Corrections to Lamb Shift and Hyperfine
Splitting}

\author {Michael I. Eides}\email{eides@pa.uky.edu,
eides@thd.pnpi.spb.ru}
\affiliation{Department of Physics and Astronomy, University of Kentucky,
Lexington, KY 40506, USA,}
\affiliation{Petersburg Nuclear Physics Institute,
Gatchina, St.Petersburg 188300, Russia}
\author{Valery A. Shelyuto}\email{ shelyuto@vniim.ru}
\affiliation{D. I.  Mendeleev Institute of Metrology,
St.Petersburg 190005, Russia}


\begin{abstract}
We consider three-loop radiative corrections to the Lamb shift
and hyperfine splitting. Corrections of order $\alpha^3(Z\alpha)^5m$
are the largest still unknown contributions to the Lamb shift in
hydrogen. We calculated radiative corrections to the Lamb shift and
hyperfine splitting generated by the  diagrams with insertions of one
radiative photon and electron polarization loops in the graphs with two
external photons. We also obtained corrections generated by the
gauge invariant sets of diagrams with two reducible radiative photon
insertions in the electron line and polarization operator insertion in
one of the radiative photons, and diagrams with two reducible radiative
photon insertions in the electron line and polarization operator
insertion  in one the external photons. Corrections  to the Lamb shift
and  hyperfine splitting generated by the diagrams with insertions of
the three-loop one-particle reducible diagrams with radiative photons
in the electron line are calculated in the Yennie gauge.
\end{abstract}



\maketitle

\section{Introduction}

The largest still unknown contributions to the Lamb shift in hydrogen
are corrections of order $\alpha^3(Z\alpha)^5m$. The magnitude of these
corrections can be easily estimated multiplying the corrections of
order $\alpha^2(Z\alpha)^5m$ (see, e.g., [1,2]) by an extra factor
$\alpha/\pi$. The contributions of order $\alpha^3(Z\alpha)^5m$ turn
out be as large as 1 kHz for the $1S$ state in hydrogen. Below we
present our recent results for three-loop radiative corrections to the
Lamb shift and hyperfine splitting.

\begin{figure}[ht]
\includegraphics[height=1.5cm]{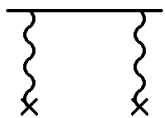}
\caption{\label{skel}
Skeleton two-photon diagram}
\end{figure}

Nonrecoil corrections of order $\alpha^3(Z\alpha)^5m$ to the Lamb shift
and corrections of order $\alpha^3(Z\alpha)E_F$ to hyperfine splitting
are generated by three-loop radiative insertions in the skeleton
diagram in Fig.\ \ref{skel}. Respective corrections of lower orders in
$\alpha$ generated by one- and two-loop radiative insertions are already
well known (see, e.g., review \cite{egs01r}). The crucial observation,
which greatly facilitates further calculations, is that the scattering
approximation is adequate for calculation  of all corrections
of order $\alpha^n(Z\alpha)^5m$ and $\alpha^n(Z\alpha)E_F$
(see, e.g., a detailed proof in \cite{eksann1}). One may easily
understand the physical reasons which  lead to this conclusion.
Consider the matrix elements of the skeleton diagram in Fig.\
\ref{skel} with the on shell external electron lines calculated
between the free electron spinors, and multiplied by the square of the
Schr{\"o}dinger-Coulomb wave function at the origin. They are described
by the infrared divergent integral

\beq        \label{nonrecskel}
-\frac{16(Z\alpha)^5}{\pi
n^3}\left(\frac{m_r}{m}\right)^3\:m\int_0^\infty\frac{dk}{k^4}
\:\delta_{l0},
\eeq

\noindent
in the case of the Lamb shift, and by the infrared divergent
integral\footnote {We define the Fermi energy $E_F$ as
\beq      
E_{F}=\frac{16}{3}Z^4\alpha^2
\frac{m}{M}(1+a_\mu)\left(\frac{m_r}{m}\right)^{3}ch\:R_{\infty},
\eeq
\noindent
where $m$ is the electron mass, $M$ is the muon mass,  $m_r$ is the
reduced mass, $\alpha$ is the fine structure constant, $c$ is the
velocity of light, $h$ is the Planck constant, $R_{\infty}$ is the
Rydberg constant, $a_\mu$ is the muon anomalous magnetic moment, and $Z$
is the nucleus charge in terms of the electron charge ($Z=1$ for
hydrogen and muonium).}

\beq        \label{skelhfs}
\frac{8Z\alpha}{\pi n^3}E_F\int_0^\infty \frac{d{ k}}{{k}^2},
\eeq

\noindent
in the case of hyperfine splitting. In these integrals
$k$ is the dimensionless momentum of the exchanged photons
measured in the units of the electron mass.

\noindent
Let us consider radiative insertions in the skeleton two-photon
diagram in Fig.\ \ref{skel}. Account of these corrections effectively
leads to insertion of an additional factor $L(k)$ in the divergent
integrals above, and while this factor has at most a logarithmic
asymptotic behavior at large momenta and does not spoil the ultraviolet
convergence of the integrals, in the low-momentum region it behaves as
$L(k)\sim k^2$ (again up to logarithmic factors), and improves the
low-frequency behavior of the integrand.  However, the integral for the
Lamb shift is sometimes still divergent after inclusion of the
radiative corrections because the two-photon-exchange diagram, even
with radiative corrections, contains a contribution of the previous
order in $Z\alpha$. This spurious contribution should be removed by
subtracting the leading low-momentum term from $L(k)/k^4$.  The result
of such subtraction is a convergent integral, where  the low
integration momenta (of atomic order $mZ\alpha$) in the exchange loops
are  suppressed,  and the effective loop integration momenta are of
order $m$. Then it is clear that small virtuality of the external
electron lines would lead to an additional suppression of the matrix
element under consideration, and it is sufficient to consider the
diagrams only with on-mass-shell external momenta for calculation of
the contributions to the energy shifts. As an additional bonus of this
approach one does not need to worry about the ultraviolet divergence of
the one-loop radiative corrections.  The subtraction automatically
eliminates any ultraviolet divergent terms and the result is both
ultraviolet and infrared finite.

\section{Diagrams with Three One-loop Electron Vacuum Polarizations}

\subsection{Lamb Shift}

\begin{figure}[ht]
\includegraphics[height=1.3cm,width=4.5cm]{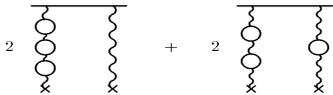}
\caption{\label{3oneloop}Three one-loop polarizations}
\end{figure}

Each polarization loop in the diagrams in Fig.\ \ref{3oneloop}
corresponds to insertion of the vacuum polarization operator
$(\alpha/\pi)k^2I_{1e}$ \cite{es03} in the Lamb shift skeleton integral
in \eq{nonrecskel}, and the contribution to the  Lamb shift generated
by the diagrams in Fig.\ \ref{3oneloop} has the form \cite{es03}

\beq
\delta E^{(1)}_{L}=-\frac{64\alpha^3(Z\alpha)^5}{\pi^4
n^3}\left(\frac{m_r}{m}\right)^3\:m\int_0^\infty{dk}{k^2}I_{1e}^3
=-~0.~021~458~(1)~\frac{\alpha^3(Z\alpha)^5}{\pi^2
n^3}\left(\frac{m_r}{m}\right)^3\:m.
\eeq

\noindent
Numerically for the $1S$ level in hydrogen we obtain $\delta
E^{(1)}_{L}=-0.002~16~\mbox{kHz}$.

\subsection{Hyperfine Splitting}

We obtain the expression for the radiative correction to hyperfine
splitting generated by the diagrams in Fig.\ \ref{3oneloop} inserting
the polarization loops in the skeleton integral in \eq{skelhfs} \cite{es03}

\beq
\delta E^{(1)}_{HFS}=\frac{32\alpha^3(Z\alpha)}{\pi^4
n^3}E_F\int_0^\infty d{ k}{k}^4I_{1e}^3
=~2.~568~3~(4)
\frac{\alpha^3(Z\alpha)}{\pi^2}\,E_F.
\eeq

\noindent
Numerically for the ground state in muonium the correction is $\delta
E^{(1)}_{HFS} =0.~003~29~\mbox{kHz}$.

\section{Diagrams with Two-Loop and  One-loop Electron Vacuum
Polarizations}

\subsection{Lamb Shift}

\begin{figure}[ht]
\includegraphics[height=1.3cm]{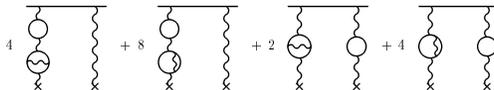}
\caption{\label{12oneloop}One- and two-loop polarizations}
\end{figure}

The integral for the diagrams in Fig.\ \ref{12oneloop}  is obtained
from the skeleton integral in \eq{nonrecskel} by insertion of the
one-loop vacuum polarization $(\alpha/\pi)k^2I_{1e}$, and the two-loop
vacuum polarization $(\alpha/\pi)^2k^2I_{2e}$ \cite{es03}

\beq
\delta E^{(2)}_{L}=-\frac{96\alpha^3(Z\alpha)^5}{\pi^4
n^3}\left(\frac{m_r}{m}\right)^3\:m\int_0^\infty{dk}I_{1e}I_{2e}
=-~0.390~152~(7)~\frac{\alpha^3(Z\alpha)^5}{\pi^2
n^3}\left(\frac{m_r}{m}\right)^3\:m.
\eeq

\noindent
Numerically for the $1S$ level in hydrogen we obtain $\delta
E^{(2)}_{L}=-0.039~21~\mbox{kHz}$.

\subsection{Hyperfine Splitting}

The expression for the hyperfine splitting generated by the diagrams in
Fig.\ \ref{12oneloop} has the form \cite{es03}

\beq
\delta E^{(2)}_{HFS}=\frac{48\alpha^3(Z\alpha)}{\pi^4
n^3}E_F\int_0^\infty d{ k}{k}^2I_{1e}I_{2e}
=~3.~559~9~(2)~
\frac{\alpha^3(Z\alpha)}{\pi^2}E_F.
\eeq

\noindent
Numerically for the ground state in muonium we obtain $\delta
E^{(2)}_{HFS} =0.004~56~\mbox{kHz}$.

\section{Diagrams with Three-Loop Electron Vacuum
Polarization}

\subsection{Lamb Shift}

\begin{figure}[ht]
\includegraphics[height=1.3cm]{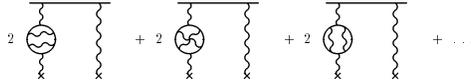}
\caption{\label{threeloop}Three-loop polarizations}
\end{figure}

For calculation of the  correction generated by the diagrams in Fig.\
\ref{threeloop} we need the three-loop vacuum polarization
operator $(\alpha/\pi)^3k^2I_{3e}$. Seven leading terms both in the
low- and  high-momentum asymptotic expansions of $I_{3e}$ over the
powers of the momentum were calculated analytically in
\cite{bb95,baik96,cks96,chks97,chks97_2}. We adjusted these results
for the case of the momentum renormalization scheme used in QED, and
constructed an interpolation which approximates the three-loop
polarization operator for all Euclidean momenta.

The skeleton integral in \eq{nonrecskel} remains infrared  divergent
even after insertion of the three-loop vacuum polarization since
$I_{3e}(0)\neq 0$. This linear infrared divergence is effectively cut
off at the characteristic atomic scale $mZ\alpha$ if we restore finite
virtualities of the external electron lines. As mentioned in
the Introduction, such infrared divergence lowers the power
of the factor $Z\alpha$, and respective would be divergent contribution
turns out to be of order $\alpha^3(Z\alpha)^4$. We carry
out the subtraction of the leading low-frequency asymptote of the
polarization operator insertion, which corresponds to the subtraction
of the leading low-frequency asymtote in the integrand for the
contribution to the energy shift ${\tilde I}_{3e}(k)\equiv
I_{3e}(k)-I_{3e}(0)$, and insert the subtracted expression in the
formula for the Lamb shift in \eq{nonrecskel}. Then the contribution to
the energy shift has the form \cite{es03}

\beq       \label{lamb3}
\delta E^{(3)}_{L}=-\frac{32\alpha^3(Z\alpha)^5}{\pi^4
n^3}\left(\frac{m_r}{m}\right)^3\:m\int_0^\infty\frac{dk}{k^2}{\tilde
I}_{3e}
=~1.015~88~(5)~\frac{\alpha^3(Z\alpha)^5}{\pi^2
n^3}\left(\frac{m_r}{m}\right)^3m.
\eeq

\noindent
Numerically for the $1S$ level in hydrogen we obtain $\delta
E^{(3)}_{L}=0.102~10~\mbox{kHz}$.

\subsection{Hyperfine Splitting}

There is no problem of infrared divergence for the radiative correction
to hyperfine splitting generated by the three-loop polarization
insertions in Fig.\ \ref{threeloop} \cite{es03}

\beq  \label{hfs3}
\delta E^{(3)}_{HFS}=\frac{16\alpha^3(Z\alpha)}{\pi^4
n^3}E_F\int_0^\infty d{ k}I_{3e}
=~1.647~9~(5)~
\frac{\alpha^3(Z\alpha)}{\pi^2}E_F.
\eeq

\noindent
Numerically for the ground state in muonium we obtain $\delta
E^{(3)}_{HFS} =0.002~11~\mbox{kHz}$.

\section{Diagrams with One-Loop Electron Factor and  Two One-loop
Electron Vacuum Polarizations}

\subsection{Lamb Shift}

\begin{figure}[ht]
\includegraphics[height=1.3cm]{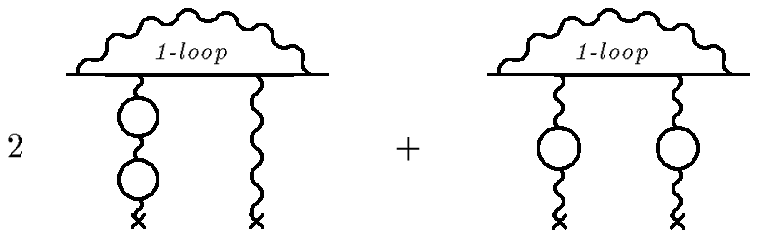}
\caption{\label{oneloopfact2looppol}One-loop electron factor and two
one-loop polarizations}
\end{figure}

\begin{figure}[ht]
\includegraphics[height=0.8cm]{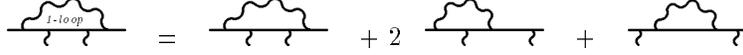}
\caption{\label{oneloopelfact}One-loop electron factor}
\end{figure}

To calculate the correction of order $\alpha^3(Z\alpha)^5$ generated by
the gauge invariant set of diagrams in Fig.\ \ref{oneloopfact2looppol}
we need a new element, namely, the gauge invariant electron factor
$L_L(k)$ in Fig.\ \ref{oneloopelfact} which describes all possible
insertions of the radiative photon in the electron line with two
external photons. An explicit expression for this electron factor was
obtained in different forms in \cite{bg87,eg,eg4,egs}. Inserting in the
skeleton integral in \eq{nonrecskel} the electron factor
$(\alpha/\pi)k^2L_L(k)$, one-loop polarization operator squared and the
multiplicity factor 3 we obtain the radiative correction in the form

\beq
\delta E^{(4)}_{L}=-\frac{48\alpha^3(Z\alpha)^5}{\pi^4
n^3}\left(\frac{m_r}{m}\right)^3\:m\int_0^\infty{dk}k^2L_L(k)I_{1e}^2.
\eeq

\noindent
It is easy to check explicitly that this integral is both ultraviolet
and infrared finite. After numerical calculations we obtain \cite{es03}

\beq  \label{lamb4}
\delta E^{(4)}_{L}=~0.0773~(4)~\frac{\alpha^3(Z\alpha)^5}{\pi^2
n^3}\left(\frac{m_r}{m}\right)^3\:m.
\eeq

\noindent
For the $1S$ level in hydrogen this contribution is $\delta
E^{(4)}_{L}=0.007~77~(4)~\mbox{kHz}$.

\subsection{Hyperfine Splitting}

We calculate the contribution to hyperfine splitting generated
by the diagrams in Fig.\ \ref{oneloopfact2looppol} using an explicit
expression for the electron factor $L_{HFS}$ like in the case of the
Lamb shift above. This electron factor was obtained in \cite{eks1}.
Inserting the electron factor $(\alpha/\pi)k^2L_{HFS}(k)$  in the
skeleton integral in \eq{skelhfs} we obtain the radiative correction in
the form \cite{es03}

\beq
\delta E^{(4)}_{HFS}=\frac{24\alpha^3(Z\alpha)}{\pi^4
n^3}E_F\int_0^\infty d{k}k^4L_{HFS}(k)I_{1e}^2
=~-3.487~2~(2)~\frac{\alpha^3(Z\alpha)}{\pi^2}E_F.
\eeq

\noindent
Numerically for the ground state in muonium this contribution is
$\delta E^{(4)}_{HFS} =-0.004~47~\mbox{kHz}$.

\section{Diagrams with One-Loop Electron Factor and  Two-Loop
Electron Vacuum Polarization}

\subsection{Lamb Shift}

\begin{figure}[ht]
\includegraphics[height=1.3cm]{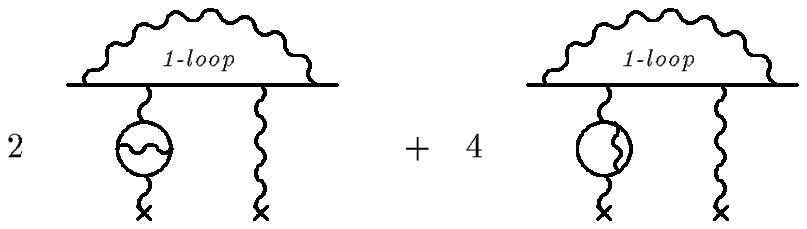}
\caption{\label{oneloopft2looppol}One-loop electron factor and
two-loop polarization}
\end{figure}

An integral representation for the correction generated by the
diagrams in Fig.\ \ref{oneloopft2looppol} is obtained from the
skeleton integral in \eq{nonrecskel}  in the standard way \cite{es03}

\beq  \label{1loopel21looppol}
\delta E^{(5)}_{L}  ~~=~~-\frac{32\alpha^3(Z\alpha)^5}{\pi^4
n^3}\left(\frac{m_r}{m}\right)^3\:m\int_0^\infty{dk}L_L(k)I_{2e}
=~ 2.191~3~(4)~\frac{\alpha^3(Z\alpha)^5}{\pi^2
n^3}\left(\frac{m_r}{m}\right)^3m.
\eeq

\noindent
Numerically for the $1S$ level in hydrogen this contribution is
$\delta E^{(5)}_{L}=0.220~24~(4)~\mbox{kHz}$.

\subsection{Hyperfine Splitting}

The radiative correction to hyperfine splitting generated by
the diagrams in Fig.\ \ref{oneloopft2looppol} has the form \cite{es03}

\beq    \label{hfs5}
\delta E^{(5)}_{HFS}=\frac{16\alpha^3(Z\alpha)}{\pi^4
n^3}E_F\int_0^\infty d{k}k^2L_{HFS}(k)I_{2e}
=~-4.680~9~(1)~
\frac{\alpha^3(Z\alpha)}{\pi^2}E_F.
\eeq

\noindent
Numerically for the ground state in muonium we obtain
$\delta E^{(5)}_{HFS} =-0.006~00~\mbox{kHz}$.

\section{Diagrams with One-Loop Polarization Insertions in the
Electron Factor and in the Externalained  Photon}

\subsection{Lamb Shift}

\begin{figure}[ht]
\includegraphics[height=1.3cm]{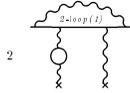}
\caption{\label{oneloopinsertelfpol}One-loop polarization insertions
in the electron factor and external photon}
\end{figure}

\begin{figure}[ht]
\includegraphics[height=0.9cm]{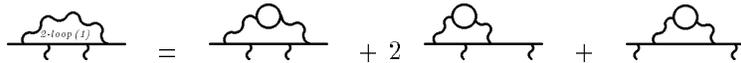}
\caption{\label{elfonelooppolins}One-loop polarization insertions in
the electron factor}
\end{figure}

The contribution to the Lamb shift generated by the diagrams in Fig.\
\ref{oneloopinsertelfpol} is similar to the contribution in
\eq{1loopel21looppol}, the only difference is that now we consider a
radiatively corrected electron factor in Fig.\ \ref{elfonelooppolins}
and a one-loop polarization insertion in the external photon.
Insertions in the skeleton integral in \eq{skel} lead to the expression
\cite{es03}

\beq  \label{lamb6gen}
\delta E^{(6)}_{L}=-\frac{32\alpha^3(Z\alpha)^5}{\pi^4
n^3}\left(\frac{m_r}{m}\right)^3\:m\int_0^\infty{dk}
L^{(2,1)}_L(k)I_{1e}
=~0.037~36~(1)\frac{\alpha^3(Z\alpha)^5}{\pi^2
n^3}\left(\frac{m_r}{m}\right)^3m,
\eeq

\noindent
where the the  electron factor $L^{(2,1)}$ with one-loop polarization
insertion in Fig.\ \ref{elfonelooppolins} was obtained in \cite{eg4}.
Numerically for the $1S$ level in hydrogen this contribution is
$\delta E^{(6)}_{L}=0.003~75~\mbox{kHz}$.

\subsection{Hyperfine Splitting}

We obtain the radiative correction to hyperfine splitting generated by
the diagrams in Fig.\ \ref{oneloopinsertelfpol} inserting the
radiatively corrected electron factor
$(\alpha/\pi)k^2L^{(2,1)}_{HFS}(k)$ \cite{eks90} in Fig.\
\ref{elfonelooppolins} in the skeleton integral in \eq{skelhfs} \cite{es03}

\beq
\delta E^{(6)}_{HFS}=\frac{16\alpha^3(Z\alpha)}{\pi^4
n^3}E_F\int_0^\infty d{k}k^2L^{(2,1)}_{HFS}(k)I_{1e}
=-0.533~3~(5)~\frac{\alpha^3(Z\alpha)}{\pi^2}E_F.
\eeq

\noindent
Numerically for the ground state in muonium this contribution is
$\delta E^{(6)}_{HFS} =-0.000~68~\mbox{kHz}$.

\section{Diagrams with Two One-Loop Polarization  Insertions in the
Electron Factor}

\subsection{Lamb Shift}

\begin{figure}[ht]
\includegraphics[height=1.3cm]{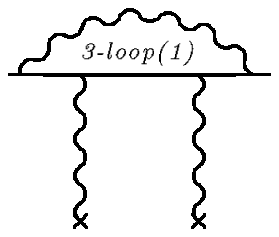}
\caption{\label{threoneloopinsertelfpol}One-loop polarization
insertions in the electron factor}
\end{figure}

\begin{figure}[ht]
\includegraphics[height=0.9cm]{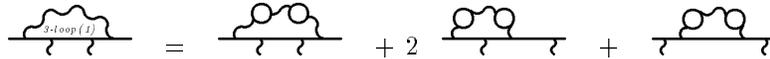}
\caption{\label{twoelfonelooppolins}Two one-loop polarization
insertions in the electron factor}
\end{figure}

The contribution to the Lamb shift generated by the diagrams in Fig.\
\ref{threoneloopinsertelfpol} is similar to the correction generated by
the one-loop polarization insertion in the electron factor calculated
in \cite{eg4}. The explicit expression for this correction contains
the electron factor with two one-loop polarization insertions
$L^{(3,1)}_L(k)$ in Fig.\ \ref{twoelfonelooppolins} \cite{es03}

\beq  \label{twopolelf}
\delta E^{(7)}_{L}=-\frac{16\alpha^3(Z\alpha)^5}{\pi^4
n^3}\left(\frac{m_r}{m}\right)^3\:m\int_0^\infty{dk}\frac{L^{(3,1)}_L(k)
-L^{(3,1)}_L(0)}{k^2}
=-0.012~610~(3)\frac{\alpha^3(Z\alpha)^5}{\pi^2
n^3}\left(\frac{m_r}{m}\right)^3m.
\eeq

\noindent
Numerically for the $1S$ level in hydrogen we obtain
$\delta E^{(7)}_{L}=-0.001~27~\mbox{kHz}$.

\subsection{Hyperfine Splitting}

The contribution to hyperfine splitting generated by the diagrams in
Fig.\ \ref{threoneloopinsertelfpol} is similar to the correction
generated by the one-loop polarization insertion in the electron factor
which was calculated in \cite{eks90}. The explicit expression for this
correction has the form \cite{es03}

\beq
\delta E^{(7)}_{HFS}=\frac{8\alpha^3(Z\alpha)}{\pi^4
n^3}E_F\int_0^\infty d{k}L^{(3,1)}_{HFS}(k)
= -0.309~05 ~(7)~\frac{\alpha^3(Z\alpha)}{\pi^2}E_F.
\eeq

\noindent
Numerically for the ground state in muonium we obtain
$\delta E^{(7)}_{HFS} =-0.000~40~\mbox{kHz}$.

\section{Diagrams with Two-Loop Polarization  Insertion in the
Electron Factor}

\subsection{Lamb Shift}

\begin{figure}[ht]
\includegraphics[height=1.3cm]{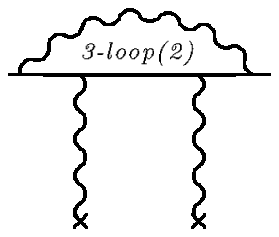}
\caption{\label{twoloopinsertelfpol}Two-loop polarization insertions
in the electron factor}
\end{figure}

\begin{figure}[ht]
\includegraphics[height=2cm]{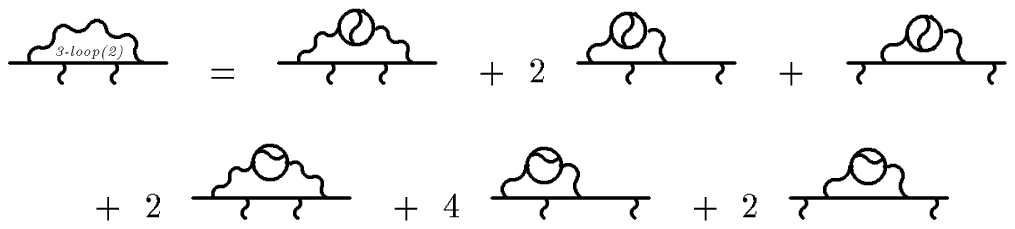}
\caption{\label{twoelflooppolins}Two-loop polarization insertions in
the electron factor}
\end{figure}

The contribution to the Lamb shift generated by the diagrams in Fig.\
\ref{twoloopinsertelfpol}  is similar to the correction generated by
the one-loop polarization insertion in the electron factor which was
calculated in \cite{eg4}. The explicit expression for this correction

\beq
\delta E^{(8)}_{L}=-\frac{16\alpha^3(Z\alpha)^5}{\pi^4
n^3}\left(\frac{m_r}{m}\right)^3\:m\int_0^\infty{dk}\frac{L^{(3,2)}_L(k)
-L^{(3,2)}_L(0)}{k^2},
\eeq

\noindent
differs from the respective expression in \cite{eg4} only due
to the difference between the electron factor with one-loop
polarization insertion $L^{(2,1)}_L(k)$ in Fig.\
\ref{elfonelooppolins} and the electron factor with
the two-loop polarization insertion $L^{(3,2)}_L(k)$ in Fig.\
\ref{twoelflooppolins}. After numerical calculations we obtain
\cite{es03}

\beq    \label{lamb8}
\delta E^{(8)}_{L} =-0.245~71~(7)\frac{\alpha^3(Z\alpha)^5}{\pi^2
n^3}\left(\frac{m_r}{m}\right)^3m.
\eeq

\noindent
For the $1S$ level in hydrogen the correction is
$\delta E^{(8)}_{L}=-0.024~70~\mbox{kHz}$.

\subsection{Hyperfine Splitting}

The contribution to hyperfine splitting generated by the diagrams in
Fig.\ \ref{twoloopinsertelfpol}  is similar to the correction generated
by the one-loop polarization insertion in the electron factor which was
calculated in \cite{eks90}. The explicit expression for this correction
has the form \cite{es03}

\beq
\delta E^{(8)}_{HFS}=\frac{8\alpha^3(Z\alpha)}{\pi^4
n^3}E_F\int_0^\infty d{k}L^{(3,2)}_{HFS}(k)
=-0.123~9~(6)~\frac{\alpha^3(Z\alpha)}{\pi^2}E_F.
\eeq

\noindent
Numerically for the ground state in muonium we obtain
$\delta E^{(8)}_{HFS} =-0.000~16~\mbox{kHz}$.

\section{Diagrams with Factorized Radiative Photons and Polarization
Operator Insertions in one the Radiative  Photons}

\subsection{Lamb shift }

\begin{figure}[ht]
\includegraphics[height=1.3cm]{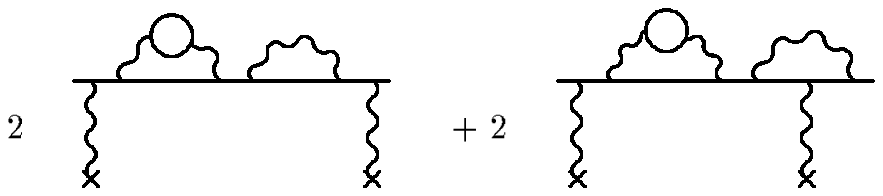}\includegraphics[height=1.3cm]{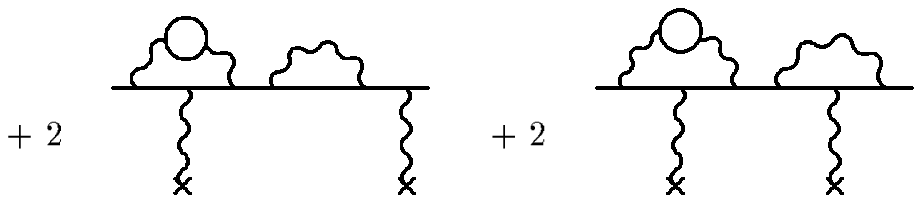}
\caption{\label{frppolrad}One-loop polarization
insertions in one of the radiative photons}
\end{figure}

We use representations

\beq
\Lambda_\mu(k, \tau)=
\frac{\alpha}{2\pi}\left\{A_{\tau}{\bf k}^2\gamma_\mu
+ B_{\tau}\gamma_\mu(\hat p-\hat k-m)
+ C_{\tau}p_\mu (\hat p-\hat k-m)
+ E_{\tau}\sigma_{\mu\lambda}k^\lambda\right\},
\eeq
\[
\Sigma (p-k, \tau)=\frac{\alpha}{2\pi}(\hat p-\hat k-m)^2
\left[M_{\tau 1}+(\hat p-\hat k) M_{\tau 2}\right],
\]

\noindent
for one-loop vertex and mass operator with massive photon, and similar
representations  $\Lambda_\mu(k)$ and $\Sigma (p-k)$ for  one-loop
vertex and mass operator with massless photon to obtain an
explicit expression for the contribution to the Lamb shift generated by
the diagrams in Fig.\ \ref{frppolrad}

\[
\delta E_{L}^{(9)}=\frac{4\alpha^3(Z\alpha)^5}{\pi^4 n^3}
\left(\frac{m_r}{m}\right)^3m
\int_0^1dv \frac{v^2\left(1-\frac{v^2}{3}\right)}{1-v^2}
\int dX\int_0^\infty dk
[(-A+B+C-E+M_1)(A_{\tau}-M_{\tau 2})
\]
\[
+(-A_{\tau}+B_{\tau}+C_{\tau}-E_{\tau}+M_{\tau 1})
(A-M_2)]
=-0.139~97~(1)~\frac{\alpha^3(Z\alpha)^5}{\pi^2 n^3}
\left(\frac{m_r}{m}\right)^3 m.
\]

\noindent
Numerically for the $1S$ level in hydrogen this correction is
$\delta E^{(9)}_{L}=-0.014~07~\mbox{kHz}$.

\subsection{Hyperfine Splitting }

The contribution to hyperfine splitting generated by the diagrams in
Fig.\ \ref{frppolrad} has the form

\beq
\delta E_{HFS}^{(9)}=
\frac{2\alpha^3(Z\alpha)}{\pi^4 n^3}~ E_F
\int_0^1dv \frac{v^2 \left(1-\frac{v^2}{3}\right)}{1-v^2}
\int {dX} \int_0^\infty {dk}
[2{\bf k}^2 AA_{\tau}-2(A E_{\tau} + A_{\tau}E)
\eeq
\[
-2(BB_{\tau}-BE_{\tau}-EB_{\tau}+EE_{\tau})
+ C(E_{\tau}-B_{\tau})
+ C_{\tau}(E-B)-2{\bf k}^2 A M_{\tau 2}
+(2B+C)(-M_{\tau 1}+M_{\tau 2})
\]
\[
+ 2EM_{\tau 1}-2{\bf k}^2 A_{\tau}M_{2}
+(2B_{\tau}+C_{\tau})(-M_{1}+M_{2})
+ 2E_{\tau}M_{1}-2(M_{1}-M_{2})(M_{\tau 1}-M_{\tau 2})
+2 {\bf k}^2 M_{2}M_{\tau 2}].
\]
\[
=0.051~44~(4)~\frac{\alpha^3(Z\alpha)}{\pi^2 n^3}E_F.
\]

\noindent
Numerically for the ground state in muonium we obtain
$\delta E^{(9)}_{HFS}=0.000~07~\mbox{kHz}$.

\section{Diagrams with Factorized Radiative Photons and Polarization
Operator Insertions in one the External Photons}

\subsection{Lamb shift }

\begin{figure}[ht]
\includegraphics[height=1.3cm]{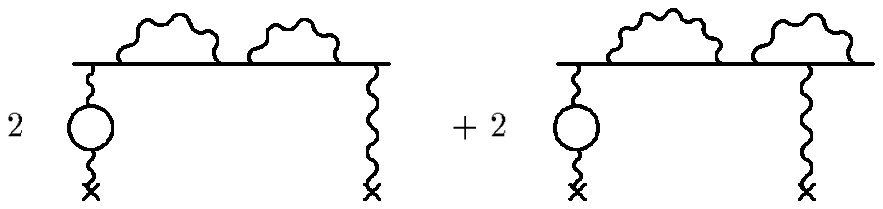}\includegraphics[height=1.3cm]{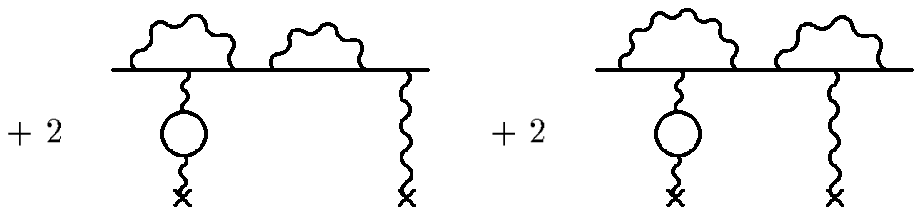}
\caption{\label{frppolext}One-loop polarization
insertions in one of the external photons}
\end{figure}

The contribution to the Lamb shift generated by the diagrams in Fig.\
\ref{frppolext} in terms of one-loop radiative corrections has the form

\[
\delta E_{L}^{(10)}=\frac{4\alpha^3(Z\alpha)^5}{\pi^4 n^3}
\left(\frac{m_r}{m}\right)^3m
\int_0^1dv\int_0^\infty dk
\frac{k^2v^2\left(1-\frac{v^2}{3}\right)}{4+k^2(1-v^2)}
\int dX[(-A+B+C-E+M_1)(A-M_2)
\]
\[
+(-A+B+C-E+M_1)(A-M_2)]
=-0.~006~25~(1)~\frac{\alpha^3(Z\alpha)^5}{\pi^2
n^3}m.
\]

\noindent
Numerically for the $1S$ level in hydrogen we obtain
$\delta E^{(10)}_{L}=-0.000~63~\mbox{kHz}$.

\subsection{Hyperfine Splitting }

The contribution to hyperfine splitting generated by the diagrams in
Fig.\ \ref{frppolext} has the form

\beq
\delta E_{HFS}^{(10)}=
\frac{2\alpha^3(Z\alpha)}{\pi^4 n^3}~ E_F
\int_0^1dv\int_0^\infty {dk}
\frac{k^2v^2 \left(1-\frac{v^2}{3}\right)}{4+k^2(1-v^2)}
\int {dX}
[2{\bf k}^2 A^2-4AE
\eeq
\[
-2(B^2-2BE+E^2)
+ 2C(E-B)
-2{\bf k}^2 A M_2
+(2B+C)(-M_1+M_2)
\]
\[
+ 2EM_1-2{\bf k}^2 AM_{2}
+(2B+C)(-M_{1}+M_{2})
+ 2EM_{1}-2(M_{1}-M_{2})^2
+2 {\bf k}^2 M_{2}^2]
\]
\[
= 0.024~8~(5)~\frac{\alpha^3(Z\alpha)}{\pi^2 n^3}E_F.
\]

\noindent
Numerically for the ground state in muonium we obtain
$\delta E^{(10)}_{HFS} =0.000~03~\mbox{kHz}$.

\section{Discussion of Results}

We calculated three-loop radiative corrections to the Lamb shift of
order $\alpha^3(Z\alpha)^5m$ and three-loop radiative corrections to
hyperfine splitting of order $\alpha^3(Z\alpha)E_F$ generated by the
gauge invariant sets of diagrams in Figs.\ \ref{3oneloop},
\ref{12oneloop}, \ref{threeloop}, \ref{oneloopfact2looppol},
\ref{oneloopft2looppol}, \ref{oneloopinsertelfpol},
\ref{threoneloopinsertelfpol}, \ref{twoloopinsertelfpol},
\ref{frppolrad}, and \ref{frppolext}. Collecting all contributions
above we obtain

\begin{equation}
\delta E_{L}=\sum_{i=1}^{i=10}\delta E_{L}^{(i)}
=2.5057~\frac{\alpha^3(Z\alpha)^5}{\pi^2 n^3} m.
\end{equation}

\noindent
Numerically for the $1S$ level in hydrogen this contribution is
$\delta E_{L}=0.252~\mbox{kHz}$.

For hyperfine splitting we obtain

\begin{equation}
\delta E_{HFS}=\sum_{i=1}^{i=10}\delta E_{HFS}^{(i)}=
-1.282~\frac{\alpha^3(Z\alpha)}{\pi^2 n^3} E_F.
\end{equation}

\noindent
Numerically for the ground state in muonium the contribution is
$\delta E_{HFS}=-0.001~7~\mbox{kHz}$.

\begin{figure}[ht]
\includegraphics[height=1.3cm]{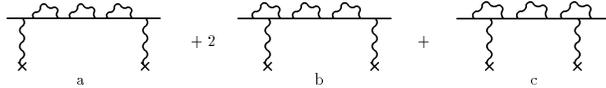}
\caption{\label{red}Reducible three-loop diagrams}
\end{figure}

We also calculated corrections of order
$\alpha^3(Z\alpha)^5m$ to the Lamb shift and corrections of
order $\alpha^3(Z\alpha)E_F$ to hyperfine splitting generated by the
diagrams in Fig.\ \ref{red}. These factorized contributions are not
gauge invariant and calculations we made in the Yennie gauge. We
obtained \cite{es04}

\beq
\Delta E_{L}=-5.~321~93~(1)~\frac{\alpha^3(Z\alpha)^5}{\pi^2
n^3}\left(\frac{m_r}{m}\right)^3\:m,
\eeq

\noindent
or $\Delta E_{L}=-0.~535~\mbox{kHz}$ for the $1S$ level in hydrogen.

Respective contribution to hyperfine splitting is \cite{es04}

\beq     \label{tothfs}
\Delta E_{HFS}  =0.~104~23~(1)~
\frac{\alpha^3(Z\alpha)}{\pi^2}\,E_F,
\eeq

\noindent
or $\Delta E_{HFS} =0.~000~13~~\mbox{kHz}$ for the ground state in
muonium.

Work on calculation of the remaining three-loop contributions to the
Lamb shift and hyperfine splitting is now in progress.

\vskip0.5cm

{\bf Acknowledgments}

\vskip0.5cm

This work was supported in part by the NSF grant PHY-0456462.  The work
of V.  A. Shelyuto was also supported in part by the RFBR grants
06-02-16156 and 06-02-04018, and by the DFG grant GZ 436 RUS 113/769/0-2.

\end{document}